\documentclass{mn2e}
\usepackage{times}
\input{psfig.sty}

\begin{document}

%\draft              

%%%%% AUTHORS -- PLACE YOUR OWN MACROS HERE %%%%%

\def\be{\begin{equation}}
\def\ee{\end{equation}}
\def\mdot{$\dot{m}$ }
\def\kms{\,{\rm {km\, s^{-1}}}}
\def\msun{{$M_{\odot}$~}}
\def\eV{{\rm \ eV}}
\def\ledd{$L_{\rm Edd}$~}
\def\mic{$\mu$ }
\def\se1{s$^{-1}$ }

\def\arcmin{\hbox{$^\prime$}}
\def\arcsec{\hbox{$^{\prime\prime}$}}
\def\degree{$^{\circ}$} 
\def\mic{$\mu$ }
\def\cm2{cm$^2$ }
\def\se1{s$^{-1}$ }

\def\gtsima{$\; \buildrel > \over \sim \;$}
\def\ltsima{$\; \buildrel < \over \sim \;$}
\def\prosima{$\; \buildrel \propto \over \sim \;$}
\def\gsim{\lower.5ex\hbox{\gtsima}}
\def\lsim{\lower.5ex\hbox{\ltsima}}
\def\simgt{\lower.5ex\hbox{\gtsima}}
\def\simlt{\lower.5ex\hbox{\ltsima}}
\def\simpr{\lower.5ex\hbox{\prosima}}
\def\la{\lsim}
\def\ga{\gsim}
\def\Lsun{\L_\odot}
\def\sr{A0620$-$00~}
\def\cxo{{\it Chandra~}}
\def\nh{N$_{\rm H}$}

\def\ie{{i.e.~}}
\def\eg{{\frenchspacing\it e.g. }}
\def\etal{{et al.~}}
%**********************************
\title[Radio outflow of A0620$-$00] 
{A radio-emitting outflow in the quiescent state of A0620$-$00: implications for modelling
low-luminosity black hole binaries}
\author[Gallo et al. ] {E. Gallo$^{1,2}$\thanks{elena@physics.ucsb.edu},
R.\,P. Fender$^{3,4}$, J.\,C.\,A. Miller-Jones$^{4}$, A. Merloni$^{5}$, 
P.\,G. Jonker$^{6,7,8}$\newauthor S. Heinz$^{9,2}$, T.\,J. Maccarone$^{3}$, M. van der Klis$^{4}$
\\ \\
$^{1}$ Department of Physics, University of California, Santa Barbara, CA 93106, USA\\     
$^{2}$ Chandra Fellow\\
$^{3}$ School of Physics and Astronomy, University of Southampton,
Highfield, SO17 1BJ, UK\\
$^{4}$ Astronomical Institute `Anton Pannekoek', University of Amsterdam,
Kruislaan 403, 1098 SJ Amsterdam, The Netherlands\\
$^{5}$ Max-Planck-Institut f\"ur Astrophysik, Karl-Schwarzschild-Strasse 1,
D-85741, Garching, Germany \\
$^{6}$ National Institute for Space Research, Sorbonnelaan 2, 3584 CA,
Utrecht, The Netherlands\\
$^{7}$ Harvard-Smithsonian Center for Astrophysics, 60 Garden Street, Cambridge, MA 02138, USA \\
$^{8}$ Astronomical Institute, Utrecht University, PO Box 80000, 3508 TA,
Utrecht, The Netherlands\\ 
$^{9}$ Kavli Institute for Astrophysics Space Research, Massachusetts
Institute of Technology, 77 Massachusetts Avenue, Cambridge, MA 02139, USA\\
}   
\maketitle
%***************
\begin{abstract} 
Deep observations with the Very Large Array of A0620--00, performed in
2005 August, resulted in the first detection of radio emission from a
black hole binary at X-ray luminosities as low as $10^{-8.5}$ times
the Eddington limit. The measured radio flux density, of $51\pm7$
$\mu$Jy at 8.5 GHz, is the lowest reported for an X-ray binary system
so far, and is interpreted in terms of partially self-absorbed
synchrotron emission from outflowing plasma.  Making use of the
estimated outer accretion rate of \sr in quiescence, we demonstrate
that the outflow kinetic power must be energetically comparable to the
total accretion power associated with such rate, if it was to reach
the black hole with the standard radiative efficiency of 10 per cent.
This favours a model for quiescence in which a radiatively inefficient
outflow accounts for a sizable fraction of the missing energy, and, in
turn, substantially affects the overall dynamics of the accretion
flow. Simultaneous observations in the X-ray band, with {\it Chandra},
confirm the validity of a non-linear radio/X-ray correlation for hard
state black hole binaries down to low quiescent luminosities, thereby
contradicting some theoretical expectations.  Taking the mass term
into account, the \sr data lie on the extrapolation of the so called
Fundamental Plane of black hole activity, which has thus been extended
by more than 2 orders of magnitude in radio and X-ray luminosity.
With the addition of the \sr point, the plane relation provides an
empirical proof for the scale-invariance of the jet-accretion coupling
in accreting black holes over the entire parameter space observable
with current instrumentation.
\end{abstract}   
%***************
%*********************
\section{introduction}
Accretion is widely recognised as the power source for the most luminous
sources in the universe; yet only for a relatively
narrow range of accretion rates, the energy release by viscous dissipation can
be radiated away efficiently by a thin inflow of plasma. While 
there is little doubt that the thin disc model (cf. Shakura
\& Sunyaev 1973) captures the basic physical
properties of black hole binaries (BHBs) in their thermal dominant (or
high/soft) state (see Davis \etal 2005), the accretion mode responsible for the
lower-luminosity (low/)hard and quiescent states is still a matter of debate
(we refer the reader to Homan \& Belloni 2005 and McClintock
\& Remillard 2006 for thorough reviews, albeit with author-dependent jargon,
on X-ray 
states of BHBs).  Since their rediscovery in recent years, radiatively
inefficient accretion flows (RIAFs) have been regarded as viable
solutions. For very low or very high accretion rates the gas cooling
efficiency can drop to a point where most of the dissipated energy is not
radiated.  Advection-dominated accretion flows (ADAFs; Ichimaru 1977; Narayan
\& Yi 1994, 1995) are popular analytical models for the dynamics of RIAFs.  At
low accretion rates, the plasma density is low enough to inhibit the Coulomb
coupling between electrons and ions, such that a significant fraction of the
viscously dissipated energy remains locked up in the gas as heat, and is
advected inward. In the case of a BH accretor, with no physical surface, such
energy would be added to the BH mass as it crossed the event horizon. Under
the ADAF working-hypothesis, the relative dimness of quiescent BHBs with
respect to quiescent neutron star X-ray binaries has been interpreted as
observational evidence for the existence of an horizon in BHBs (Narayan
\& Yi 1995;  Narayan, McClintock \& Yi, 1997; Narayan, Garcia, \& McClintock
1997; Garcia \etal 2001; 
McClintock, Narayan \& Rybicki 2004; but see Jonker \etal 2006).  It has been
noted that ADAFs may be convectively unstable under certain viscosity ranges,
hence the notion of convection-dominated accretion flows (CDAFs; Quataert \&
Gruzinov 2000; Narayan, Igumenshchev, Abramowicz 2000), where the inward
angular momentum transport by convective motions cancels the outward transport
by viscous torques, yielding nearly zero net accretion rate.  Blandford \&
Begelman (1999; 2004) argue that the accreting gas in an ADAF is generically
unbound and free to escape to infinity, and elaborate an alternative model,
named adiabatic inflow outflow solution (ADIOS). Here the key notion is that
the excess energy and angular momentum is lost to a wind at all radii; the
final accretion rate into the hole may be only a tiny fraction of the mass
supply at large radii, leading to a much smaller luminosity than would be
observed from an efficient inflow. Another possible scenario for
low-luminosity BHs is that proposed by Merloni \& Fabian (2002), where strong,
unbound, magnetic coronae are powered by thin discs, with the relative
fraction of the power liberated in the corona increasing as the accretion rate
decreases. 
Models in which the X-ray emission would be
generated by the coronae of the secondary stars (Bildsten \& Rutledge 2000)
seem to be inconsistent with the observed spectral shapes and luminosities of
quiescent BHB systems (e.g. Kong \etal 2002; Hameury \etal 2003; McClintock
\etal 2003).

Regardless of which model provides the correct description for the
inflow of gas, in recent years it has become clear that a second
component must be taken into account in order to reproduce the
broadband spectral energy distribution (SED) of low-luminosity BHBs in
the hard and quiescent states (where the--somewhat arbitrary--boundary
between the two states can be set around $10^{33.5}$ erg sec$^{-1}$;
McClintock \& Remillard 2006). At least down to Eddington-scaled X-ray
luminosities of a few $10^{-6}$, these systems display flat or
slightly inverted radio-mm spectra (Fender 2006 and references
therein), interpreted as synchrotron emission from a continuously
replenished, partially self-absorbed collimated outflow (cf. Blandford
\& K\"onigl 1979; Hjellming \& Johnston 1988; Kaiser 2006). Such jets
may even contribute to the X-ray power-law emission of hard state BHBs
by means of optically thin synchrotron and synchrotron self-Compton
radiation from the innermost region (Falcke \& Biermann 1995; Markoff,
Falcke \& Fender 2001; Markoff \& Nowak 2004; Markoff, Nowak \& Wilms
2005; Giannios 2005).  Corbel \etal (2003) and Gallo, Fender \& Pooley
(2003; GFP03 hereafter) have established a quantitative coupling
between accretion and the production of jets in hard state BHBs, in
terms of a tight correlation between the X-ray and the radio
luminosity ($L_{\rm X}$ and $L_{\rm R}$), of the form $L_{\rm
R}\propto L_{\rm X}^{0.7\pm 0.1}$. The correlation extends over more
than 3 orders of magnitude in $L_{\rm X}$ and breaks down around 2 per
cent of the Eddington X-ray luminosity $L_{\rm Edd}$, above which the
sources enter the thermal dominant state, and the core radio emission
drops below detectable levels.  Probably the most notable implication
of this non-linear scaling is the predicted existence of a critical
X-ray luminosity below which a significant fraction of the liberated
accretion power is channelled into a radiatively inefficient outflow,
rather than being dissipated locally by the inflow of gas and emitted
in the form of X-rays (this does not necessarily imply the X-ray
spectrum is dominated by non-thermal emission from the jet, as most of
the jet power may be stored as kinetic energy). Conservative estimates
indicate that such a threshold luminosity should be no lower than
$\sim 4\times 10^{-5} L_{\rm Edd}$, thereby encompassing the whole
quiescent regime (Fender, Gallo \& Jonker 2003). The same radio/X-ray
scaling found for BHBs holds for super-massive black holes in active
galactic nuclei (AGN) when a mass term $M$ is included in the
analysis: Merloni, Heinz, Di Matteo (2003; MHdM03 hereafter) and
Falcke, K\"ording \& Markoff (2004;FKM04), have independently proven
that accreting BHs (the binaries plus a sample of some 100 AGN) form a
fundamental plane (FP) in the $\rm log$($L_{\rm R}, L_{\rm X}, M$)
domain (see also Bregman 2005 and Merloni \etal 2006). The FP, for
AGN, extends to Eddington ratios as low as $ 10^{-11}$, where the very
existence of BHB accretion remains to be proven. In addition, it has
been argued that the radio/X-ray correlation might break down
somewhere below $10^{-5} L_{\rm Edd}$ (MHdM03; Heinz 2004; Yuan \& Cui
2005).  In light of these issues, and given the sensitivity
limitations of current radio telescopes, deep simultaneous radio and
X-ray observations of nearby quiescent systems are needed in order to
explore the process of jet formation (if any) in this regime, test the
underlying accretion flow models, probe and extend the FP beyond its
current limits.

{\sr} (=V616\,Mon) was discovered in outburst in 1975 August (Elvis \etal
1975); for about two months, it was the brightest X-ray source in the sky.
During the onset of the outburst, the radio counterpart was first detected at
a level of 80 mJy at 2.4 GHz (Craft 1975; see Davis \etal 1975 for a previous
upper limit). Subsequent observations revealed highly variable radio emission,
with a peak flux density of 300 mJy at 1.4 GHz (Owen \etal 1976).
Kuulkers \etal (1999) collected all the available radio data for the 1975
outburst (see their Table 1 for a full list of references in chronological
order) and found evidence of multiple 
(at least three) ejection events with expansion velocities in excess of
$0.5c$.   
Within 15 months from the outburst peak, \sr had faded back to its quiescent regime, in which
it has remained ever since.  
In 1986 March, McClintock \& Molnar (see
McClintock, Horne \& Remillard 1995) established an upper limit to the radio
counterpart in quiescence of 0.14 mJy at 4.8 GHz.  Several infrared/optical/UV studies
undertaken over the past three decades have determined the system parameters
to a high level of accuracy (see Shahbaz \etal 2004 and references
therein). Most important for this study, \sr lies at a distance of $1.2\pm
0.4$ kpc (Shahbaz, Naylor \& Charles 1994; Gelino \etal 2001; Jonker
\& Nelemans 2004) and has a dynamically established BH accretor 
with mass $M=11.0\pm1.9~M_{\odot}$ (Shahbaz \etal 1994;
Gelino et al. 2001).
X-ray observations of the system in quiescence, performed with
{\it ROSAT} (McClintock \etal 1995; Narayan, Barret \& McClintock 1997), {\it ASCA} (Asai
\etal 1998) and
\cxo (Kong \etal 2002) have revealed a variable (by a factor of about 2) X-ray
source with an average 0.4-2.4 keV 
luminosity of $\sim 3\times 10^{30}$ erg sec$^{-1}$, corresponding to an
Eddington-scaled luminosity of a few $10^{-9}$.
Due to its low luminosity and relative proximity,
\sr represents the most suitable known system to probe 
the FP beyond its current range.

In this Paper we report on deep radio observations of \sr performed in
2005 August with the Very Large Array (VLA), observed simultaneously 
in the X-ray band with {\it Chandra}, and discuss the implications of
their outcomes for modelling low-power accreting BHs.

%**********************
\section{Data analysis}
%*************************************
\subsection{Chandra X-ray observatory}

\sr was observed with the back-illuminated S3 chip of the ACIS detector
on board \cxo on 2005 August 20, starting at Universal Time (UT) 08:36 (MJD
53602.3589), and for $\sim$39.6 ksec (ObsId 5479).  Standard pipeline-processed level 2
data were employed for the analysis, performed using the {Chandra Interactive
Analysis of Observations} ({\sc CIAO}) software, version 3.2.1.

\sr was detected at right ascension (R.A.) 06$^{\rm h}$ 22$^{\rm m}$ 44$^{\rm s}$.54 and
declination (Dec.) $-$00$^{\circ}$ 20$^{\prime}$ 44.38$^{\prime\prime}$
(J2000), consistent with the optical position given by Liu, van Paradijs \&
van den Heuvel (2001). We first searched for background flares by inspecting
the 0.3-10.0 keV light curve of the S3 chip, having first subtracted the
regions containing sources (8 sources were identified by the {\it celldetect}
tool). Only intervals with count rates between 0.7 and 1.2 count sec$^{-1}$
(note that the background typically climbs by a factor of 8 between 7-10 keV)
were retained for the analysis, resulting in a loss of 0.3 ksec worth of data.
Unless otherwise noted, all quoted uncertainties are given
at the 90 per cent confidence level. 
 
We extracted data from a circle of 10 pixels in radius centred on A0620$-$00,
and background from an annulus with inner and outer radii of 10 and 18 pixels;
316 counts were detected in the source region (covering an area of $\sim$314
pixels), and 30 counts in the background annulus (covering an area of
$\sim$704 pixels), yielding a net time-averaged rate of $7.7 \pm 0.5
\times 10^{-3}$ count sec$^{-1}$ after background subtraction.  We then fitted
the source light curve with a constant function,
and obtained a reduced $\chi^2$ of 0.6, with 13 degrees of freedom (d.o.f.),
indicating that no statistically significant variability took place during the
observation.
The radial profile of the source and that of the normalised Point Spread
Function are consistent with each other within the errors, indicating that
\sr is a point-like source in the
\cxo image.

\begin{figure}
\psfig{figure=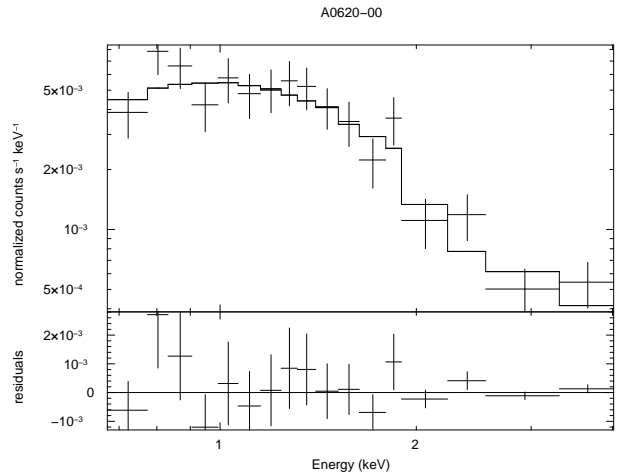,width=0.5\textwidth,angle=-90}
\caption{\label{spectrum}{\em Top:} X-ray spectrum of \sr as observed by \cxo
on 2005 
August 20. The spectrum is well
fitted by an absorbed power law model with photon index $\Gamma=2.1$ and
equivalent hydrogen column density $N_{\rm H}= 1.6\times 10^{21}$ cm$^{-2}$
(see Table\ref{tab_spec}). The corresponding 2.0-10.0 keV source flux,
corrected for absorption, is $4.1\times 10^{-14}$ erg cm$^{-2}$ sec$^{-1}$.
{\em Bottom:} residuals after subtracting the best-fitting model from the
data. }
\end{figure}

Using the same source and background regions as for the count rate analysis,
we extracted the spectrum of \sr over the energy interval 0.3-8.0 keV using
the CIAO tool {\it psextract} (the ancillary response file was automatically
corrected for the time-variable low-energy quantum efficiency degradation of
the detector). The spectrum was grouped into energy bins containing at
least 15 counts and analysed with XSPEC, version 11.2.0.
An absorbed power law with photon index $\Gamma=2.08^{+0.49}_{-0.35}$ provides a
good fit to the data.  The fitted value of the equivalent hydrogen column
density is consistent, within the errors, with the optical value of
$1.94\times 10^{21}$ cm$^{-2}$ (Predehl
\& Schmitt 1995), whereas the values obtained by fitting the spectrum with
thermal models, such as a bremsstrahlung or blackbody, are significantly lower
(in fact they hit the minimum values allowed by the fitting program). The
best-fitting model obtained by letting the hydrogen column density free to
vary, i.e. the power law model, is shown in Figure \ref{spectrum},
over-plotted to the 0.3-8.0 keV \cxo spectrum; the parameters are listed in
Table
\ref{tab_spec}. 

The estimated 2.0-10.0 keV unabsorbed flux is $F_{\rm
  X}=4.1^{+0.5}_{-1.4} \times 10^{-14}$ erg cm$^{-2}$ sec$^{-1}$ (the
flux error algorithm within XSPEC was set to draw 1000 sets of
parameter values from the fitted distribution; the quoted flux
uncertainties are given at the 68 per cent confidence level; for
reference, the flux over the 0.5-10.0 keV band is
$6.7^{+0.8}_{-2.3}\times 10^{-14}$ erg cm$^{-2}$ sec$^{-1}$). Adopting
a distance of $1.2\pm0.4$ kpc, the resulting X-ray luminosity over the
2--10 keV energy interval is $L_{\rm X} = 7.1 _{-4.1} ^{+3.4} \times
10^{30}$ erg sec$^{-1}$, where the distance and spectral shape
uncertainties contribute similarly to the errors\footnote{The quoted
  uncertainties have been estimated by applying the standard error
  propagation formulae, being aware that the large fractional errors
  on both distance and flux formally invalidate the assumptions of
  small errors behind such formulae. The same caveat applies to the
  radio luminosity estimate.  }.  This corresponds to $\sim 5 \times
10^{-9}L_{\rm Edd}$ for a 11 $M_{\odot}$ BH.

The fitted spectral slope is consistent with the results from a previous \cxo
observation, performed in 2000 February. Kong \etal (2002), applying both CASH
statistics and $\chi^2$ statistics to the data, found $\Gamma=2.19\pm0.50$, whereas 
McClintock \etal (2003) obtained a somewhat softer value
($\Gamma=2.26\pm0.18$) by grouping the data into 10 spectral bins by 12 counts and
applying $\chi^2$ statistics.
Previous observations with {\it ROSAT} (Narayan
\etal 1997), gave $\Gamma=3.5\pm0.7$, with $N_{\rm H}$ fixed to the optical
value. For completeness, we re-analysed the 2000 \cxo observation of
\sr (ObsId 95) and fitted the sum of the 2000 and 2005 spectra.
The extraction of the 2000 spectrum was performed following the same steps as
described above; following Kong \etal (2002), only intervals for which the
source-free count rate was lower than 0.15 count sec$^{-1}$ were examined.
The combined 0.3-8.0 keV spectrum, grouped into bins by at least 15 counts, is well
fitted by an absorbed power law with photon index
$\Gamma=2.06^{+0.18}_{-0.25}$ and hydrogen column density $N_{\rm
H}=2.04^{+0.65}_{-0.79}\times 10^{21}$ cm$^{-2}$. The inferred 2.0--10.0 keV
unabsorbed flux is $2.7^{+0.2}_{-0.5}\times 10^{-14}$ erg cm$^{-2}$
sec$^{-1}$ (68 per cent confidence level), corresponding to an average
2.0-10.0 keV luminosity $L_{\rm X}=4.7\pm0.4 \times 10^{30}$ erg sec$^{-1}$. 

%---------------------------------------------------------------------
\begin{table*}
\caption{\label{tab_spec} Best-fitting spectral parameters of \sr as
  observed by \cxo on 2005 August 20. All quoted
  uncertainties are at the 90 per cent confidence
level. The goodness 
  of the fit is expressed by the reduced $\chi^2$ for a certain number of 
  degrees of freedom (d.o.f.).}  \centering
\begin{tabular}{lcccc}
\hline
Model      & $N_{\rm H}$   & $\Gamma$ & kT & $\chi^2_{\rm red}$/d.o.f. \\
           & (10$^{21}$ cm$^{-2}$)     & & (keV)&         \\
\hline
Power law  &  1.62$^{+1.51}_{-1.36}$ & 2.08$^{+0.49}_{-0.35}$&...& 0.83/14 \\

           &  1.94 (fixed)           & 2.17$\pm 0.27$&...& 0.80/15 \\
Bremsstrahlung & 0.64$^{+1.33}_{-0.64}$ &...& 3.46$^{+6.62}_{-1.65}$& 0.88/14\\
           & 1.94 (fixed) &...&2.06$^{+1.09}_{-0.59}$ & 0.99/15 \\
Blackbody  & 0.00$^{+0.05}$ &...& 0.46$\pm$0.06 & 1.68/14\\
           & 1.94 (fixed) &...& 0.39$^{\star}$& 2.39/15 \\
\hline
\end{tabular}
\flushleft $^{\star}$ No error is calculated for reduced $\chi^2$ higher than
2.
\end{table*}

%****************************
\subsection{Very Large Array}
\label{sec:radio_data}
A0620$-$00 was observed with the VLA in its C-configuration at a
frequency of 8.46\,GHz.  Three sets of observations were made between
2005 August 19 10:23 UT and 2005 August 20 21:12 UT (MJD
53601.4326--53602.8833), giving a total of 855\,min on A0620$-$00
itself. The \cxo observation started about 6 ksec before the last 
VLA set, meaning that for the remaining 33.6 ksec the X-ray/radio coverage
was strictly simultaneous.  

VLA Data calibration and imaging were performed using standard procedures
within the National Radio Astronomy Observatory's (NRAO) {\sc Astronomical
Image Processing System (aips)}.  3C\,48 was used as the primary calibrator,
using the flux density scale derived at the VLA in 1999 as implemented in the
31Dec05 version of AIPS.  The secondary calibrator was J\,0641$-$0320, at an
angular separation of 6.03$^{\circ}$ from the target source, with an 8.4-GHz
flux density of $0.639 \pm 0.001$ Jy.  15-min observations of A\,0620$-$00
were interleaved with 90-sec observations of the phase calibrator.  The
observations were made using standard procedures, with two IF pairs of width
50\,MHz centred at 8.435 and 8.485\,GHz.

\begin{figure}
\vspace{0.3cm}
\begin{center}
\psfig{figure=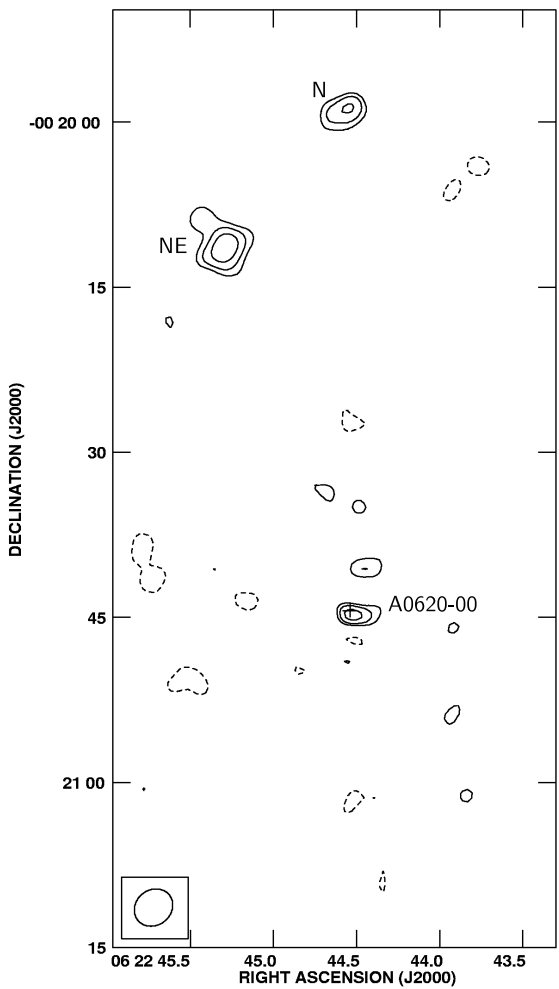,width=0.45\textwidth,angle=0}
\caption{Radio contour map of the field surrounding A0620$-$00 as observed by
the VLA on 2005 August 19-20; the \cxo position of \sr is marked with a cross,
whose size corresponds to the error on the X-ray position.  The beam size is
$3.7^{\prime\prime}\times3.2^{\prime\prime}$, and the rms noise in the image is
7.0\,$\mu$Jy\,bm$^{-1}$.  Solid and dashed contours are at $\sqrt{2}^n$ and
$-(\sqrt{2}^n)$ times the lowest contour level of 20\,$\mu$Jy\,bm$^{-1}$.  In
addition to A0620$-$00, we detect two other sources to the north (Table 2).}
\label{fig:radio}
\end{center}
\end{figure}

Figure \ref{fig:radio} shows a naturally weighted contour map of the observed
field.  An unresolved (down to a beam size of $3.7^{\prime\prime}\times
3.2^{\prime\prime}$) radio source is visible at a position consistent with
A0620$-$00, whose X-ray position is marked with a cross. The fitted radio
source position is R.A.  06$^{\rm h}$22$^{\rm m}$44.503$^{\rm
s}\pm$0.012$^{\rm s}$ Dec. $-$00$^{\rm
d}$20$^{\prime}$44.72$^{\prime\prime}\pm$0.10$^{\prime\prime}$ (J2000). The
source is detected at a flux density $S_{\rm 8.46~GHz}=51.1$
$\mu$Jy\,beam$^{-1}$, a $7.3\sigma$ detection at the rms noise level of
7.0\,$\mu$Jy\,bm$^{-1}$; the quoted noise level is a factor 1.4 higher than
the theoretical thermal noise limit, possibly because the source was too weak
to use self-calibration.

In order to quantify the probability that the detected radio flux is
extragalactic in origin, we made reference to the catalogue of published
source counts at 1.4~GHz in the literature, complete down to 50 $\mu$Jy (Huynh
et al. 2005). Integrating their polynomial fit to the differential source
counts from a minimum value (see below) up to `infinity' (that is 1000 Jy for
practical purposes), gives the number of counts expected per square arcsec.
Using a minimum flux density of 147 $\mu$Jy, which is the 1.4~GHz flux density
corresponding to 50 $\mu$Jy for a non-thermal spectrum typical for
extragalactic radio sources, where the monochromatic flux density, $S_{\nu}$,
scales as $\nu^{\alpha}$, with $\alpha=-0.6$, gives a probability of $6 \times
10^{-5}$ of having an extragalactic background source within the same distance
of the X-ray position as our radio source. This effectively rules out an
extragalactic origin for the detected emission.

At a distance of 1.2$\pm$0.4 kpc, the measured flux 
corresponds to a radio luminosity $L_{\rm R}=7.5 \pm 3.7 \times 10^{26}$ erg
sec$^{-1}$, with the quoted error bar being dominated by the distance
uncertainty.
In approximating the integrated radio luminosity as the monochromatic
luminosity multiplied by the observing frequency, we have assumed a flat radio
spectrum of spectral index $\alpha=0$, as is usually observed in the hard
X-ray state of BHBs, with a spectrum extending up to $\nu_{\rm max}=8.46$\,GHz
(see Section 3 for details on the interpretation of the radio
emission). The integration is performed between this upper bound and minimum
frequency $\nu_{\rm min}\ll\nu_{\rm max}$.  However, it is important to
mention that several hard state sources, such as XTE J1118+480 (Fender
\etal 2001), GX339--4 (Corbel \etal 2000), XTE J1550--564 (Corbel \etal 2001),
actually display inverted radio-mm spectra, with $\alpha\simeq +0.5$, or even
higher, up to $+0.7$ over some epochs. If this was the case in \sr, this would
introduce a further source of error in the luminosity estimate, although
negligible with respect to the distance uncertainties.

The two bright sources to the north and northeast of A0620$-$00 are of
similar brightness to the source itself.  The parameters of all three
sources are given in Table \ref{tab_obs}.  
The two northern sources
are separated from one another by 16.4$^{\prime\prime}$.  
The 2MASS catalogue shows no obvious match to the northern sources, 
and no central
source between the two (as might be expected if they were associated
with the lobes of a radio galaxy).  
If they are extragalactic, the
possibility arises of using them to measure the proper motion of
A0620$-$00.

\begin{table*}
\begin{center}
\caption{\label{tab_obs} Parameters of the three radio sources in Figure 4.}
\begin{tabular}{cccc} \hline 
Source & RA (J\,2000) & Dec.\ (J\,2000) & Flux density ($\mu$Jy\,bm$^{-1}$)\\
\hline
A0620$-$00 & 06$^{\rm h}$22$^{\rm m}$44.503$^{\rm s}\pm$0.012$^{\rm s}$ &
$-$00$^{\rm d}$20$^{\prime}$44.72$^{\prime\prime}\pm$0.10$^{\prime\prime}$
& $51.1\pm6.9$\\
N source & 06$^{\rm h}$22$^{\rm m}$44.577$^{\rm s}\pm$0.018$^{\rm s}$ &
$-$00$^{\rm d}$19$^{\prime}$59.03$^{\prime\prime}\pm$0.20$^{\prime\prime}$ & $43.0\pm6.9$\\
NE source & 06$^{\rm h}$22$^{\rm m}$45.295$^{\rm s}\pm$0.014$^{\rm s}$ & $-$00$^{\rm d}$20$^{\prime}$11.44$^{\prime\prime}\pm$0.23$^{\prime\prime}$
& $52.2\pm6.7$\\
\hline 
\end{tabular}
\end{center}
\end{table*}
%----------------

%*************************************
\section{Radio emission in quiescence}

One of the main questions we wish to address in this work concerns the hard
state jet survival at low quiescent luminosities.  The deep VLA observations
presented here provide us with the first detection of radio emission from a
BHB at $L_{\rm Edd}\simeq10^{-8.5}$, 3 orders of magnitude deeper than
any detection reported so far.  The measured radio flux represents
the lowest level of radio emission, both in flux and luminosity, ever reported
for an accreting BH. Prior to these observations, the lowest Eddington-scaled
X-ray luminosities for which radio emission had been detected from a BHB were between a few
$10^{-6}$ to about $10^{-5}$, respectively in V404 Cyg and GX~339--4 (Gallo
\etal 2005; Corbel \etal 2003). Since neutron star X-ray binaries tend to be
dimmer in the radio band (Migliari \& Fender 2006), the measured radio flux is
actually the lowest reported for a Galactic X-ray binary system.
We point out that the radio detection of A0620--00, which is among the
dimmest BHBs detected in X-rays so far (e.g. McClintock \etal 2003; Tomsick
\etal 2003), really pushes the performances of current radio telescopes to the
limit. For comparison, GS~2000+25, whose quiescent X-ray luminosity is a
factor of about 2 lower than \sr (Garcia
\etal 2001) is located at a distance of 2.7 kpc (Jonker \& Nelemans 2004 and
references therein). If its radio luminosity was comparable to that of
A0620--00, it would take
$8\times12$~hr on source for the VLA in order to get a
5$\sigma$ detection. The next generation of radio arrays, such as e-MERLIN and
EVLA, will be able to achieve this sensitivity in about 7 hr.

%
%Radio emission from X-ray binaries is generally interpreted as synchrotron
%radiation from some kind of outflow 
As far as the interpretation is concerned, the synchrotron nature of the radio
emission from X-ray binaries is generally inferred from the high brightness
temperatures, high degree of polarisation and non-thermal spectra, whereas the
outflowing nature is inferred by brightness temperature arguments, combined
with the persistent level of radio emission (Hjellming \& Han 1995; Mirabel \&
Rodr\'\i guez 1999; Fender 2006).
As in the case of \sr we have none of the above-mentioned features at our
disposal, we must explore different mechanisms as well.
Fitting the infrared-to-X-ray spectrum of \sr in quiescence with a pure ADAF
model severely under-predicts the measured radio flux: the extrapolation of an
(admittedly old version of an) ADAF spectrum for \sr down to radio frequencies
underestimates the 8.5 GHz flux by more than 3 orders of magnitude (see
Narayan McClintock
\& Yi 1996).
Despite the significant improvements in the ADAF model over the past decade,
to take into account effects such as the possible energy advection by
electrons and substantial wind mass loss, it seems unlikely that the radio to
X-ray spectra of hard and quiescent states can be reproduced without invoking
an extra component.  Even though various models for the quiescent state of
BHBs predict a substantial fraction of the inflowing gas be lost to a wind,
far too high mass loss rates would be required in order to reproduce the
measured radio fluxes in terms of e.g. free-free emission (see Gallo, Fender
\& Hynes 2005, and references therein, for a more quantitative argument
applied to V404 Cyg).  

At such low luminosity levels we must also explore the possibility that
the radio emission may be contaminated by gyro-synchrotron radiation from the
corona of the companion star.  The
secondary star in \sr is a 0.7 $M_{\odot}$ K3-K4V star, in a 7.75 hr orbit
around the BH.  Being in forced synchronous rotation, it will be a very fast
rotator; the observed rotational broadening of the spectral lines of the
companion star, together with an estimate of the inclination, gives a velocity
$v\simeq 130$ km sec$^{-1}$ (Marsh, Robinson \& Wood 1994; Shahbaz \etal 1994;
Gelino, Harrison \& Orosz 2001). 
Typical radio luminosities from early-to-mid type K stars vary in the range
$1\times 10^{14}$ up to $3\times 10^{15}$ erg sec$^{-1}$ Hz$^{-1}$ in the most
extreme cases. HD 197890 (Speedy Mic), a rapidly rotating single K star, with
orbital period of 0.42 days and $v\times \rm sin (i)\simeq 170$ km sec$^{-1}$,
has a radio luminosity of $2.7\times 10^{15}$ erg sec$^{-1}$ Hz$^{-1}$ at 8.4
GHz (Robinson \etal 1994). 
AB Dor, a similar object also with period of
 about half a day, has been detected at about 5 mJy (Lim et al. 1994).
At a distance of 15 pc, this implies $L_{\rm R} \simeq 1.4\times 10^{15}$
erg sec$^{-1}$ Hz$^{-1}$.
Note that the orbital parameters have been proven
not to influence significantly the level of radio emission from magnetically
active stars (Drake, Simon \& Linsky 1989; see G\"udel 2002 for a thourough review). In
fact, radio emission from tight binary star systems is significantly weaker
than that of RS CVn binaries or magnetically active single stars, possibly
because of reduced differential rotation, weaker dynamo effects or a change in
the energy transfer mechanism.   
As the measured radio emission from \sr is a 
factor of 30 higher than the highest radio luminosities for magnetically
active stars of comparable spectral type, we conclude that such a
contamination is likely to be negligible.

Radio emission from \sr is also unlikely to be associated with a major
optically thin radio flare, as such phenomena in BHBs tend to be bright ($>$
mJy), short-lived (few hr timescales) and have always been associated with
outbursts at all wavelengths\footnote{but see Pal \& Chakrabarti (2004) who claim the
detection of a $\sim$190 sec micro-flare, with peak flux of 3.84 mJy at 1.28
GHz, during a 1 hr long observation of \sr with the Giant Meter Radio
Telescope in 2002.}.
Radio emission from radio lobes presumably resulting from the interaction of 
highly relativistic ejecta with the surrounding interstellar medium, of the
kind observed in e.g. 1E1740.7--2942 (Mirabel \etal 1992; 1993), GRS 1758--258
(Mirabel \etal 1993), Cyg X-1 (Gallo \etal 2005), are 
typically observed on much larger (arcmin) scales. 
Thus we are led to the conclusion
that the quiescent radio emission is probably generated in a continuously
replenished relativistic outflow, associated with a persistent level of radio
emission and flat or inverted radio-mm spectrum (however, this does not
necessarily rule out short time-scale variability, of the kind observed in
V404 Cyg; Hynes
\etal 2004a). As for V404 Cyg, only high sensitivity, high spatial resolution
radio observations will ultimately answer the question of whether the radio
emitting outflow is collimated at such low flux levels.

It has been suggested that the outflow, and not the inflow of gas,
could be responsible for the hard X-ray power law which dominates the
spectrum of hard state BHBs (Falcke \& Bierman 1995; Markoff \etal
2001; Giannios 2005).  In response to the criticism offered by several
authors (e.g. Zdziarski \etal 2003; Heinz 2004; Maccarone 2005), the
jet model has also been significantly improved over the past years.
In a recent work, Markoff, Nowak \& Wilms (2005) explore the
possibility that the jet `base', i.e. a hot wind blowing away from the
inner regions of the accretion flow, and is later collimated into the
actual jet, in fact coincides with the Comptonising corona itself.
Here the radio-to-soft X-rays are dominated by synchrotron emission,
while the hard X-rays are due to inverse Compton scattering at the jet
base/corona, with both disc and synchrotron photons acting as seed
photons. The authors find that the jet and Compton coronal models
describe high statistics observations of GX 339--4 and Cyg X-1 while
in the hard state equally well. At least in one source (GX~339--4),
the pure-jet model is able to reproduce analytically the slope of the
observed radio/X-ray correlation (Corbel \etal 2003) by varying the
fractional jet power (Markoff \etal 2003).  It would be interesting to
see whether the same model could account for the SED of quiescent BHB
systems as well; this will be explored in a future paper.

Yuan, Cui \& Narayan (2005) developed a coupled jet-ADAF model for the SED of
XTE J1118+480 while in the hard X-ray state (see also Meier 2001). Unlike in
the model by Markoff and collaborators, here the X-ray emission is still
dominated by a highly advective inflow, while the outflow emission is modelled
by means of multiple internal shocks in a mildly relativistic highly
collimated jet. The model also seems to account for the complex timing
behaviour and optical/X-ray time lags observed in this source (Kanbach \etal
2001; Malzac
\etal 2003). It is worth mentioning that, within this framework, the estimated
outer accretion rate is about one tenth of Eddington, comparable to the 
values inferred in the thermal 
dominant state BHs. In any case, the model generalisation (Yuan \& Cui 2005)
predicts that the observed radio/X-ray correlation in hard state BHBs (GFP03)
should break down at low quiescent luminosities. This is not observed, as
discussed in Section~\ref{sec:dom}.

%*****************************************************
\subsection{\sr and the Radio/X-ray/Mass correlations} 

The simultaneous {\it Chandra}/VLA observations of A0620--00, together with the
dynamical estimate of the BH mass in this system, allow us to test and extend
the FP of BH activity (MHdM03; FKM04) orders of magnitude beyond
its current limits. We wish to stress that the only way this can be done is to
probe the low luminosity, low-mass corner of the distribution (i.e. to the
left of the plane as viewed edge-on in Figure 4 of MHdM03). This is because the
right hand side of the relation is already occupied by the most massive BHs we
know, accreting at around the Eddington rate.

\begin{figure}
\psfig{figure=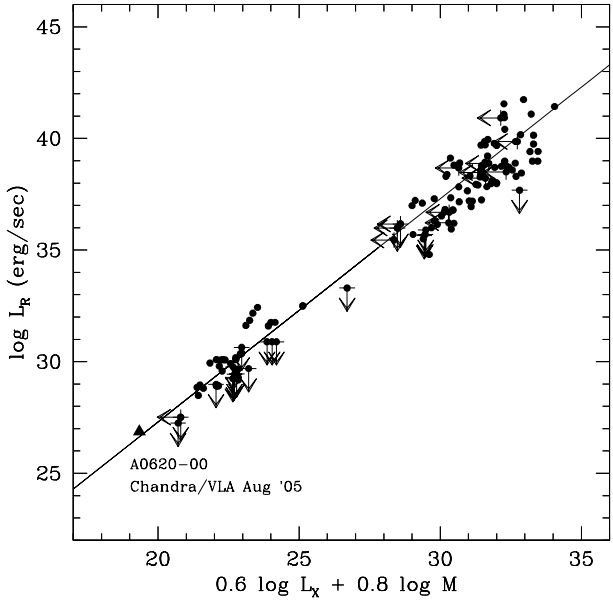,width=8.5cm,angle=0}
\caption{\label{fp} The Fundamental Plane (FP) of
black hole activity of Merloni, Heinz \& Di Matteo (2003; see also 
Falcke, K\"ording \& Markoff 2004) with the addition of
A0620$-$00. Thanks to the simultaneous {\em Chandra}/VLA detection of
this system, marked by a triangle, the FP is
extended by more than two orders of magnitude on the x-axis. 
}
\end{figure}

With the addition of the \sr data, the sample of MHdM03 (the only
genuinely beamed source, 3C 273, has been removed from the original
sample; see Merloni \etal 2006) is well fitted by the same expression
given in the original paper: $~{\rm log} L_{\rm R} = (0.6\pm0.1){\rm
  log} L_{\rm X} + (0.8 \pm 0.1) {\rm log} M + (7.3\pm4.1) $.  The
logarithmic deviation from the FP is 0.93$\sigma$, within even the
measurement error. The validity of the FP is thus extended by 2.5
orders of magnitude in $L_{\rm X}$ and 2.3 orders of magnitude in
$L_{\rm R}$.  Figure~\ref{fp} shows the FP, with the \sr point marked
by a triangle in the bottom left corner.  We note that the current
estimate of the BH mass in \sr (Gelino \etal 2001) is based on the
assumption that the accretion disc does not contaminate the IR
emission from the system, whereas recent observations show that a disc
contamination is, at least sometimes, present (Hynes, Robinson \&
Bitner 2005). This would result in a lower BH mass. For reference,
assuming 5 \msun for the mass of the BH in \sr would in result in a
$\chi^2$ compared to the mean of 0.51 (vs. 0.18 for $M=11$ \msun).

This 3D correlation has led to two different proposals for
unifying low-power BHs. While FKM04 argue that the plane
relation follows naturally from a model in which the X-ray emission from
sub-Eddington BHs (hard and quiescent systems in the case of the binaries) is
dominated by optically thin synchrotron radiation from the innermost region of
the jet 
(cf. Falcke \& Bierman 1995), MHdM03 apply the theoretical relations by Heinz
\& Sunyaev (2003), and conclude that the FP is inconsistent with the X-ray
emission coming from a radiatively efficient inflow and is only marginally
consistent with optically thin synchrotron from the jet. In a refined analysis
which includes a proper treatment of the electrons' cooling, Heinz (2004)
concludes that X-ray synchrotron emission would be as unlikely as efficient
disc emission. As discussed at length by both groups in a recent work (Merloni
\etal 2006), while the measured slope of the FP may depend on the sample
selection, the agreement between the slopes for the BHB and for the AGN sample
indicates that the slope uncertainty cannot exceed by much that quoted in the
original papers.  However, it remains true that, within those errors,
there is margin for both models (RIAF-plus-jet or pure-jet) to be accepted
(but see K\"{o}rding, Falcke \& Corbel 2006, who argue that using hard
state objects alone, the plane relation is in agreement with the prediction of a
pure jet model for the emitted radiation). 
It is worth mentioning that Roberston \& Leiter (2004) propose a radically
different explanation for the plane relation, in terms of a magnetic propeller
effect that requires intrinsically magnetic central compact objects.  
\label{sec:dom}
\begin{figure}
\psfig{figure=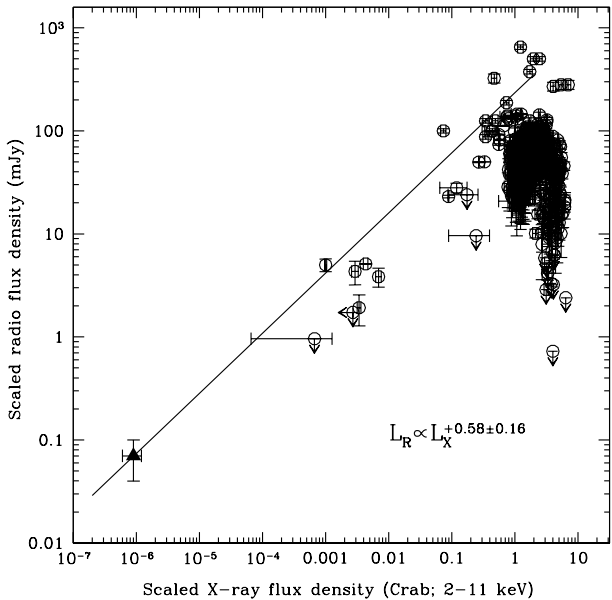,width=8.5cm,angle=0}
\caption{\label{bin} The BHB sample of Gallo, Fender \& Pooley (2003),
  with the addition of the \sr {\it Chandra}/VLA data. The hard state
  sources -- below a flux of 2.5 Crab, scaled to a distance of 1 kpc
  -- are well fitted by a non linear radio/X-ray relation of the form
  $L_{\rm R}\propto L_{\rm X}^{0.58\pm0.16}$ (solid line), implying
  that the ratio between the radio and X-ray luminosities tends to
  increase towards quiescence. The fitted slope is admittedly affected
  by the uncertainties in the distance to GX~339--4 (see text); the
  quoted value is given for D=6 kpc.}
\end{figure}

In light of these issues, while it may be still premature to use the
FP to constrain the accretion flow model, the \sr data (minus 3C 273)
-- with $\rm log$$(L_{\rm X})=30.84$, $\rm log$$(L_{\rm R})=26.87$ and
$\rm log$$(M)=1.04$ -- confirm that a scale-invariant description of
the jet-accretion coupling is supported by observations of accreting
(candidate) BHs spanning {\it fourteen} orders of magnitude in $L_{\rm
  X}$, {\it fifteen} orders of magnitude in $L_{\rm R}$, and {\it
  eight} orders of magnitude in $M$.\\

For completeness, we also re-fitted\footnote{As in GFP03, only data up
  to 2.5 Crab (scaled to 1 kpc) were fitted, with the exclusion of Cyg
  X-1. } the BHB sample of GFP03 with the addition of the \sr
data. The large uncertainties in the distance to GX 339--4 (Hynes
\etal 2004b), for which the correlation holds over 3 orders of
magnitude in $L_{\rm X}$, affect the fitted slope.  Assuming a
distance of 6 kpc to this source (instead of 4 kpc adopted in the
original paper, before Hynes \etal 2004b), results in $ L_{\rm
  R}\propto L_{\rm X}^{0.58\pm 0.16}$ (see Figure~\ref{bin}).
Adopting instead the (less likely) maximum distance of 15 kpc gives a
slope of $0.68\pm 0.03$. The fitted slopes are consistent with each
other within the errors, and with the $0.7\pm0.1$ given by GFP03. We
note however that a slope of $0.71 \pm 0.01$ was first found between
the radio and X-ray {\it fluxes} in GX339--4 (Corbel \etal 2003) over
different epochs and accretion regimes. The same slope, albeit with
larger errors, was later found for V404 Cyg (GFP03). This might
indicate a flattening of the correlation at very low luminosities. It
is worth mentioning that the low-luminosity Galactic Centre BH,
Sagittarius A$^{\ast}$, appears to lie well above the fundamental
plane when in the non-X-ray-flaring regime (see Markoff 2005).
Pending a more accurate determination of the actual slope value, a
{\it non linear} radio/X-ray scaling for hard state BHBs appears to
hold over 6.5 orders of magnitude in X-ray luminosities, between
$10^{-8.5}$ to a few $10^{-2}L_{\rm Edd}$. Above this critical
luminosity, and allowing for hysteresis effects (Maccarone \& Coppi
2003), the systems make a transition to the thermal dominant state,
and their core radio emission is quenched below detectable levels.
This latter point clearly highlights a further level of complexity,
which is not apparent when the BHB and AGN samples are blended (see
however Maccarone, Gallo \& Fender 2003, and Greene, Ho \& Ulvestad
2006, who suggest that a partial quenching of radio emission from AGN
takes place in the same Eddington-scaled luminosity interval as for
the binaries).  The criticism by Bregman \etal (2005) does not apply
to BHB systems, the argument there being that one should restrict the
study to a sample of objects for which the range in the distance
squared is less than that in $L_{\rm R}$ or $L_{\rm X}$ in order to
find meaningful relations. Not only the range in distance squared for
the binaries is less than 4 orders of magnitude smaller with respect
to that in $L_{\rm X}$, but also the correlation has been shown to
hold for two different sources over a period of several years (Corbel
\etal 2003; GFP03).  We refer the reader to Merloni \etal (2006) for a
fuller discussion on the robustness of the FP.
%*********************************
\section{Assessing the outflow contribution to the global energetics and dynamics} 

Prior to this work, two major uncertainties concerned: 1. whether
quiescent BHBs are still producing radio emitting outflows in
quiescence; 2. if so, whether the same non-linear radio/X-ray scaling
found in hard state sources holds down to very low
luminosities. Nearly simultaneous radio/X-ray observations of the BHB
XTE J1908+094 during the decay of an outburst (Jonker \etal 2004)
suggested that the radio/X-ray correlation might break down at low
luminosities -- caveat the not strict simultaneity of those
observations combined with the very fast decay rate.  However, the \sr
data imply a positive answer to both questions. If the radio/X-ray
relation does break down at some point, as has been suggested on the
basis that the synchrotron X-rays from the jet scale more slowly with
the accretion rate than the X-ray flux from the accretion flow
(e.g. Heinz 2004), this does not happen down to at least
$10^{-8.5}L_{\rm Edd}$. This appears to be at odds with the prediction
by Yuan \& Cui (2005), whose coupled ADAF-jet model predicts that the
correlation should turn and become steeper (assuming the form $L_{\rm
  R}\propto L_{\rm X}^{1.23}$) somewhere below $10^{-5}$ to
$10^{-6}L_{\rm Edd}$.

The confirmation of a non-linear radio/X-ray scaling down to low
quiescent luminosities lends observational support to the notion of
`jet-dominated states' as defined by Fender \etal (2003), i.e. a
regime in which the energy carried off by an outflow -- in the form of
radiation, kinetic energy and Poynting flux -- exceeds the X-ray power
output (notice that this is not necessarily the same proposal by
Falcke \& Bierman, 1995; 1996; 1999, where the {\it radiation} output
is jet-dominated, and the underlying inflow is even dimmer than in
e.g. the ADAF picture).  
This notion follows from the non-linear radio/X-ray scaling under the  
{\it assumption} that the total power output from hard/quiescent
state X-ray binaries is the sum of the radiative luminosity of the flow
and total jet power, i.e. that no extra `missing energy', most notably
advection, is needed in order to
reproduce the observed low X-ray luminosities. 
At the same time, while ADAF models predict the existence of 
bipolar outflows emanating from the surface layers of the equatorial inflow (see Narayan \&
Yi 1995), 
generally they do not address the importance of such outflows 
with respect to the overall accretion process in terms of energetics.

Having established the presence of a radio-emitting outflow in the
quiescent state of A0620--00, in the following we shall attempt to
test whether its contribution can be relevant to the flow energetics
and dynamics, under two representative working-hypotheses: ADAF
vs. ADIOS.  The boundary condition of the problem is set by the
accretion rate in the outer disc of \sr during quiescence, $\dot
M_{\rm out}$.
Based on models for the optical/UV emission of the outer accretion
disc in dwarf novae (Warner 1987), corrected downward to account for
the mass difference, McClintock et al. (1995) estimate $\dot M_{\rm
  out} = y 10^{-10} M_{\odot}$ yr$^{-1}$ for A0620--00, where $y$ is a
factor of the order unity, that can be up to a few.  This value is
also comparable to the $3\times 10^{-11}$ \msun yr$^{-1}$ (e.g. Huang
\& Wheeler 1989) inferred from the measure of the total energy
released during the 1975 outburst of \sr adopting 58 year recurrence
time based on plate archives which showed an outburst in 1917 (Eachus
et al. 1976). This time-averaged value had been calculated based on a
distance of 1 kpc for \sr (vs. a refined value of 1.2 kpc) and could
still be underestimated by a factor 2 or so, to allow for the
possibility that an intermediate outburst was
missed\footnote{Especially considering that that in the mid 40's,
  close to World War II, there were no useful Harvard plate
  archives.}.

The putative luminosity
associated with  $\dot M_{\rm out}$, {if it was to reach the black
  hole with standard radiative efficiency}, would be
\begin{equation}
L_{\rm tot} \equiv \eta  \dot M_{\rm out}c^2 \simeq 6 \times 10^{35}
y~(\eta / 0.1)~\rm erg ~\rm sec^{-1}~~,
\end{equation}
much larger than the observed X-ray (or bolometric) luminosity.
In the above formula $\eta$ is the accretion efficiency, which depends only on
the BH spin.
The various RIAF models provide different explanations
for the much lower luminosities that are observed in terms of different
`sinks' for the energy. 
\begin{enumerate}
\item {
In the {\bf ADAF} scenario, it is assumed that all the $\dot M_{\rm out}$ is
accreted onto the black hole ($\dot M_{\rm in}=\dot M_{\rm out}$)
while the radiated luminosity 
$L_{\rm bol}=\epsilon_{\rm rad} \dot M_{\rm in}c^2$ 
is much smaller
as a result of a reduced {radiative efficiency}
\begin{equation}
\label{eq:radeff}
\epsilon_{\rm rad} \equiv \eta f(\alpha) =\eta \times \left\{
        \begin{array}{ll}
        1,   &  \dot M_{\rm out} \ge \dot M_{\rm cr}   \\
        (\dot M_{\rm out}/\dot M_{\rm cr})^{\alpha}, & 
\dot M_{\rm out} < \dot M_{\rm cr}   \\
        \end{array}\right.\;
\end{equation}
where  
$\dot M_{\rm cr}$ is the
critical rate above which the disc becomes 
radiatively efficient. The index 
$\alpha$ is typically close to unity, but its exact value may depend on the
micro-physics of ADAF and on how the bolometric luminosity is
calculated (for example, MHdM03 found that, if
$L_{\rm bol}$ is estimated from the 2--10 keV luminosity with a
constant correction, then $\alpha\simeq 1.3$; but see Yuan \& Cui
2005 for a different scaling).
Writing the 
bolometric luminosity of \sr as $L_{\rm bol}= w~10^{32}$ erg sec$^{-1}$, where we have
introduced a multiplicative factor $w$, we have
that 
\begin{equation}
f (\alpha)= \left(\frac{\dot M_{\rm out}}{\dot M_{\rm cr}}\right)^{\alpha} \simeq 1.7
\times 
10^{-4}~(w/y)~(0.1/\eta).
\end{equation}  
For $\alpha=1.3$, 
\begin{equation}
\dot M_{\rm cr}\simeq 8\times 10^{-8}~y^{(1+1/1.3)}~w^{(-1/1.3)}~(\eta/0.1)^{(1/1.3)} ~M_{\odot}~\rm yr^{-1}
\end{equation}   
which corresponds to an Eddington-scaled critical accretion rate  
$\dot m_{\rm cr} \simeq 0.36~y^{(1+1/1.3)} w^{(-1/1.3)}
(\eta/0.1)^{(1+1/1.3)}$.  With $w\sim  
5-10$, as implied by the ADAF spectral modelling (see Figures 6 and 7 in
Narayan \etal 1997), a self-consistent ADAF solution is
obtained for $\dot m_{\rm cr}$ of a few times $10^{-2}$, as expected on
theoretical grounds. 

Contrary to the ADIOS case, for the ADAF scenario to be
self-consistent, the total kinetic power of the jet/outflow, $L_{\rm kin}$, 
should be a
negligible fraction of $L_{\rm tot}$.  In order to verify this, we can
estimate $L_{\rm kin}$ making use of the normalisation for the jet kinetic
power vs. radio luminosity 
derived by
Heinz \& Grimm (2005). Obviously, if this number is high enough, the jet
contribution to the flow energetics (and dynamics) can be negligible with
respect to advective cooling. 
In Heinz \& Grimm (2005) the radio core emission of three well studied
radio galaxies (M87, Per A and Cyg A) was directly compared to the radio lobe
emission, used a jet calorimeter. They proposed that the jet kinetic power
can be estimated from the core radio luminosity in the following way:
\begin{equation}
\label{eq:kin}
L_{\rm kin} = 6.2 \times 10^{37} \left(\frac{L_{\rm R}}{10^{30}{\rm
      erg}\;{\rm s}^{-1}}\right)^{\frac{1}{1.4-\alpha_r/3}}{\cal
      W}_{37.8}~{\rm erg}\; {\rm sec}^{-1}
\end{equation}
where $\alpha_r$ is the radio spectral index, and the parameter ${\cal
  W}_{37.8}$ carries the (quite large) uncertainty on the radio galaxy
calibration\footnote{Following the formalism of Heinz \& Grimm (2005),
  the {\it average} jet `efficiency' $W_0$ is given by the expression
  $W_0 \approx 6.2\times 10^{37}{\cal W}_{37.8}$ erg sec$^{-1}$ (see
  their Equation 10). Here the parameter ${\cal W}_{37.8}$, which
  equals unity if $W_0=10^{37.8}$ erg sec$^{-1}$, is meant to allow
  the reader to adjust for future improvements and/or preferences in
  this value.  Note however that the normalisation in
  Equation~\ref{eq:kin} is very close to that proposed by Fender,
  Maccarone \& van Kesteren 2006 based on different grounds.}.  For
\sr, assuming a flat radio spectral index $\alpha_r=0$, and for
$L_{\rm R}=7.5 \pm 3.7 \times 10^{26}$ erg sec$^{-1}$, we obtain
$L_{\rm kin} \simeq 3.6 \times 10^{35}{\cal W}_{37.8}$~erg sec$^{-1}$,
or
\begin{equation}
\frac{L_{\rm kin}}{L_{\rm tot}} \simeq  0.6 \times {\cal W}_{37.8}~{y}^{-1}~(\eta/0.1)^{-1}
\end{equation}
This would mean that the jet/outflow carries a significant amount of
the accretion energy budget away from the system. I so, then any realistic 
accretion flow model for quiescence shall necessarily incorporate the
effects of an outflow both in terms of energetics and dynamics,
effectively ruling out a pure ADAF solution.

Using the estimate for the kinetic power of the jet in quiescence
given in Equation~\ref{eq:kin}, we can calculate the total energy
carried out by it in between outbursts, assuming again that the 58
years recurrence time is not overestimated.  We obtain $E_{\rm
  jet}\simeq 6.6 \times 10^{44} {\cal W}_{37.8}$ erg, of the same
order of the energy released during an outburst. Interestingly,
Meyer-Hofmeister \& Meyer (1999) calculated the outburst evolution for
\sr with a model accounting for evaporation of the cold outer disc
(but neglecting outflows), and concluded that only about one third of
the mass accreted during quiescence need to be stored in the disc for
the subsequent outbursting episode.\\}

\item{
Under the {\bf ADIOS} working-hypothesis, the mass flowing from the outer disc does
not reach 
the inner region, but is lost in a outflow. The accretion rate is now a
function of radius:
\begin{equation}
\dot M(R)=\left\{ \begin{array}{ll}
        \dot M_{\rm out},   & R_{\rm out}>R>R_{\rm tr}   \\
        \dot M_{\rm out}(R/R_{\rm tr})^{\alpha}, & R_{\rm tr}>R>R_{\rm
        in}\\
        \end{array}\right.\;
\end{equation}
where we have assumed that mass loss sets in within the truncation
radius $R_{\rm tr}$. Proceeding as before, we can then estimate the
truncation radius for $\dot M_{\rm in}=\dot M(R_{\rm in}$=$3R_{\rm S})$,
being $R_{\rm S}$ the Schwarzschild radius for a 10 \msun BH.
For $\alpha=1$ we obtain
\begin{equation}
R_{\rm tr}\simeq 1.8\times 10^4~(y / w)~(\eta/ 0.1)~R_{\rm S} \;,
\end{equation}
or $\simeq 5.4\times 10^{10}~(y/w)~(\eta/0.1)$ cm, 
to be compared with the orbital separation of about 
$3\times 10^{11}$ cm (Gelino \etal 2001). 

If we assume that the
the jet/outflow is powered by the mass lost from the accretion flow,
then its total kinetic power $L_{\rm kin}$ is given by
\begin{equation}
L_{\rm kin}=L_{\rm tot} \left(1-\frac{R_{\rm
      in}}{R_{\rm tr}}\right)\simeq L_{\rm tot},
\end{equation} 
i.e. in the ADIOS framework, a dominant fraction of the total accretion power is
channelled into the jet/outflow.} 
\end{enumerate}

In spite of the many uncertainties in the above calculations, the
jet/outflow kinetic power turns out to be a sizable fraction of
$L_{\rm tot}$, effectively ruling out a {\em pure} ADAF solution for
the dynamics of \sr in quiescence.  

However, within these uncertainties there is still room for a hybrid
solution to apply, one in which at each $\dot{M}$ about half of the
energy is carried away by the outflow, while the rest is advected
inward and finally added to the BH mass (see e.g. K\"ording, Fender \& Migliari 
2006).

%**********************************
\section{Summary and final remarks}

Deep VLA observations of A0620--00, performed in 2005 August, have provided us
with the first radio detection of a quiescent stellar mass BH emitting at
X-ray luminosities as low as $L_{\rm Edd}\simeq 10^{-8.5}$.  The level of
radio emission -- 51 $\mu$Jy at 8.5 GHz -- is the lowest ever measured in an
X-ray binary. At a distance of 1.2 kpc, this corresponds to a radio luminosity
$L_{\rm R}=7.5\times 10^{26}$ erg sec$^{-1}$. By analogy with higher
luminosity systems, partially self-absorbed synchrotron emission from a
relativistic outflow appears to be the most likely interpretation. Free-free
wind emission is ruled out on the basis that far too high mass loss rates would
be required, either from the companion star or the accretion disc, to produce
observable emission at radio wavelengths, while gyrosynchrotron radiation from
the corona of the companion star is likely to contribute to less than 5 per
cent to the measured flux density.

\sr was observed simultaneously in the X-ray band with {\it Chandra}; the 0.3-8
keV spectrum is well fitted by an absorbed power law with photon index
$\Gamma=2.08^{+0.49}_{-0.35}$ and hydrogen equivalent column density
consistent with the optical value. The corresponding 2--10 keV luminosity is 
$L_{\rm X}=7.1\times 10^{30}$ erg sec$^{-1}$, a factor of about two higher
than in a previous \cxo observation, in February 2000. Combining the spectra
of the two \cxo observations results in a best-fitting power law photon
index $\Gamma=2.04^{+0.18}_{-0.25}$, not particularly soft with respect to the
$\Gamma\simeq 1.7$ characteristic of higher luminosity, hard state BHBs.

The simultaneous \cxo observation of \sr allowed us to test and extend the
radio/X-ray correlation for BHBs by 3 orders of magnitude in $L_{\rm
X}$.  The measured radio/X-ray fluxes confirm the existence of a non-linear
scaling between the radio and X-ray luminosity in this systems; with the
addition of the \sr point {$L_{\rm R}\propto L_{\rm X}^{0.58\pm0.16}$ provides
a good fit to the data for $L_{\rm X}$ spanning between $10^{-8.5}$ and
$10^{-2} L_{\rm Edd}$. The fitted slope, albeit consistent with the previously
reported value of $0.7\pm0.1$, is admittedly affected by the uncertainties in
the distance to GX339--4, for which the correlation extends over 3 orders
of magnitude in $L_{\rm X}$ and holds over different epochs. Pending a more
accurate determination of the distance to this source, we can nevertheless
exclude the relation breaking down and/or steepening in quiescence.  

By making use of the estimate of the outer accretion rate of \sr in
quiescence and of the jet radiative efficiency by Heinz \& Grimm
(2005), we demonstrate directly that the outflow kinetic power
accounts for a sizable fraction of the accretion energy budget, and
thus must be important with respect to the overall accretion dynamics
of the system. If indeed the outflow acts as a primary sink of energy
in quiescence, then its effects should be incorporated in the
modelling of the underlying accretion flow, favouring ADIOS-like
scenarios to pure ADAF models.  As envisaged by Livio \etal (2003), it
may well be that what is generally interpreted as a physical disc
inner radius in the hard and quiescent states, actually corresponds to
the transition between the portion of the disc where the bulk of the
liberated accretion power is dissipated locally, within the
disc-corona system, and the inner portion, where the most of the power
is channelled into a relativistic outflow. In other words, the disc
only disappears observationally. A qualitatively similar scenario is
that proposed by Ferreira \etal (2006), where the central regions of
hard/quiescent state BHBs are composed of an outer standard disc and
an inner `jet-emitting disc', driving a MHD jet. It is worth
mentioning that global, fully relativistic MHD simulations of
accretion tori also result in Poynting flux dominated narrow jets
emanating from the innermost regions, surrounded by a wider,
baryon-loaded outflow (De Villiers \& Hawley 2003).

The radio/X-ray correlation for BHBs has been proven to extend down to the
low-mass low-luminosity corner of the so called `Fundamental Plane of black
hole activity' (MHdM03; FKM04), which unites stellar and super-massive BHs in
a $\rm log$($L_{\rm R}$-$L_{\rm X}$-$M$) plane.  The \sr data lie on the best
fitting 3D correlation, which has thus been extended by more than 2 orders of
magnitude in $L_{\rm X}$ and $L_{\rm R}$. 
While it may be still premature to make use of the plane relation
as a conclusive diagnostic of the accretion model, the FP can be
read as an observational proof for the scale-invariance of the jet-accretion
coupling in accreting black holes spanning the whole range of radio and X-ray
luminosities that is observable with current instrumentation.

%*************************
\section{Acknowledgements}

Support for this work was provided by the National Aeronautics and
Space Administration (NASA) through \cxo Postdoctoral Fellowship
Awards PF5-60037 (E.G.)  and PF3-40026 (S.H.), issued by the {\it
  Chandra} X-Ray Observatory Center, which is operated by the
Smithsonian Astrophysical Observatory for and on behalf of NASA under
contract NAS8-03060. The National Radio Astronomy Observatory is a
facility of the National Science Foundation operated under cooperative
agreement by Associated Universities, Inc..  E.G. is grateful to
Manuel G\"udel for helpful suggestions on the radio properties of
magnetic stars. T.M. wishes to thank Elmar K\"ording for useful
discussions.
\vspace{-0.5cm}
%************************

\end{document}